# Non-bolometric bottleneck in electron-phonon relaxation in ultra-thin WSi film

Mariia Sidorova*
*Physics Department, Moscow State Pedagogical University, Russia
and DLR Institute of Optical Systems, Rutherfordstrasse 2, 12489 Berlin, Germany*

A. Kozorezov
*Department of Physics, Lancaster University, United Kingdom*

A. Semenov and A. Korneev
*Physics Department, Moscow State Pedagogical University, Russia
and Moscow Institute of Physics and Technology, Russia*

Yu. Korneeva
*Physics Department, Moscow State Pedagogical University, Russia*

M. Mikhailov
*B. Verkin Institute for Low Temperature Physics and Engineering of the National Academy of Sciences of Ukraine, Kharkiv, Ukraine*

A. Devizenko
*National Technical University "Kharkiv Polytechnic Institute", Ukraine*

G. Chulkova and G. Goltsman
*Physics Department, Moscow State Pedagogical University, Russia
and National Research University Higher School of Economics, Russia*

We developed the model of internal phonon bottleneck to describe the energy exchange between the acoustically soft ultra-thin metal film and acoustically rigid substrate. Discriminating phonons in the film into two groups, escaping and non-escaping, we show that electrons and non-escaping phonons may form a unified subsystem, which is cooled down only due to interactions with escaping phonons, either due to direct phonon conversion or indirect sequential interaction with an electronic system. Using an amplitude-modulated absorption of the sub-THz radiation technique, we studied electron-phonon relaxation in ultra-thin disordered films of tungsten silicide. We found an experimental proof of the internal phonon bottleneck. The experiment and simulation based on the proposed model agree well, resulting in $\tau_{e-ph}\sim$ 140-190 ps at $T_C$ = 3.4 K supporting the results of earlier measurements by independent techniques.

## I. INTRODUCTION

Since the demonstration of the first Superconducting Nanowire Single-Photon Detector (SNSPD) [1] more than a decade ago, many research groups around the world have attempted to improve SNSPD performance. SNSPDs play an important role in many applications where single-photon detection is of prime importance [2]. SNSPDs demonstrate an ultrafast timing performance, i.e. picosecond timing jitter [3] and nanosecond reset time, record detection efficiency [4], ultralow noises and wide spectral sensitivity from visible to near infrared wavelengths [5]. Among many studied superconducting materials: NbN [1], Nb [6], NbTiN [7], NbC [8], TaN [9], $MgB_2$ [10], NbSi [11], MoGe [12] and MoSi [13], the most recent advance was achieved with SNSPD based on an amorphous superconductor, tungsten silicide, WSi, that allowed the development of a device with a record detection efficiency exceeding 90% [14]. Nevertheless in spite of rapid progress, real devices fail to combine record performance metrics simultaneously. The main reason why this has not been achieved yet is the absence of a complete understanding of the detection mechanism.

The detection mechanism in SNSPD is connected with the formation of a resistive state in a current-carrying superconducting nanowire. It is triggered by the absorption of a photon. Whilst the exact origin of the resistive state was the subject of extensive recent theoretical works [14-18], the initial excited state, being no less important, was not studied in detail. The excited state is created in the energy down-conversion process following photon absorption and the initial energy deposition in the form of energetic electronic excitations. From initial electronic excitations it proceeds via an avalanche-type cascading process, characterized by a multiplication of the number of carriers and phonons with a continuous spectral downslide, until the excited volume containing highly non-equilibrium quasiparticles and phonons, termed as a hotspot, is formed [15]. The details of this process play an important role in SNSPDs and were the focus of recent work [18], where the dominant role of electron-phonon interactions over electron-electron interactions was emphasized even in strongly disordered NbN and WSi where electron-electron interaction is strongly enhanced. Another aspect of hotspot formation is phonon loss into a substrate, defining energy density and hence equilibration rates [18] and fluctuations [19]. The energy exchange between the film and the substrate controls the nucleation and growth of normal domain, resulting in photon count. The latter is also influenced by the strength of electron-phonon, phonon-electron interactions and phonon escape from the film. For this reason, studying electron-phonon interaction and cooling of non-equilibrium electron and phonon distributions in materials, which are used for radiation detection is one of the central problems for all sensors.

A new class of amorphous superconductors for SNSPDs, and in the first instance WSi, attracted immediate attention. Subsequently, electron-phonon interaction in WSi was studied by applying pump probe [20] and magnetoconductance [21] measurements, while the detection mechanism was investigated with quantum detector tomography [22]. The experimental work [20] utilised a time resolved two-photon detection technique to study the evolution of the hotspot in current-carrying nanowire under the conditions that the nanowire remains superconducting. Relaxation times of the order of hundreds of picoseconds were found and interpreted in [23] using the kinetic model of hotspot relaxation, where self-recombination of non-equilibrium quasiparticles plays a dominant role. The characteristic electron-phonon time $\tau_0$ [24], which depends on material, was found to be 0.84 -1.0 ns [20] for tungsten rich WSi. Thus, electron-phonon relaxation in WSi turned out to be slow in comparison with conventional SNSPD materials, such as NbN and NbTiN (where the measured time of relaxation of electron temperature at critical temperature, $T_C$, was of the order of 10 ps) [25]. The slower electron-phonon relaxation in WSi in the temperature range close to material critical temperature is of no surprise, being a consequence of lower $T_C$ in WSi in comparison with NbN, assuming that sound velocities are close for both materials. The exact strength of the electron-phonon interaction in WSi is an important characteristic, not only determining its superconducting properties, but arguably being significant for superior detection capabilities of this material.

Recently new experimental data were obtained for WSi applying the magnetoconductance technique [21]. In the temperature range ~10 – 20 K, the electron-phonon relaxation time was observed to scale with temperature as $\tau_{ph-e} \sim T^{-3}$ and the extrapolated value of $\tau_{ph-e}$ at $T_C=4.5$ K was 66 ps leading to $\tau_0 \approx 2.5$ ns derived under the assumption of a partial thermalisation of electrons [18]. These estimates favour the hotsot detection mechanism in WSi SNSPDs for nanowires with the width $W \geq 150$nm [18]. The value of $\tau_0 \approx 5.0$ ns was obtained in [19] finding an agreement between experimental data in WSi SNSPD of 130 nm width and the hot-belt model in contrast with the hot-spot model predictions.

Apart from the remaining uncertainties in the estimates of electron-phonon interaction strength in WSi the energy loss from the utra-thin film into the substrate was specifically emphasized in [19] as an important problem for the realization of a particular detection mechanism and resulting in fluctuations affecting the performance of SNSPD. Correspondingly, the main objective of this paper is the theoretical and experimental study of the mechanisms of electron-phonon interaction and the energy exchange between ultra-thin metal films and dielectric substrates. Experiments were performed on WSi film grown on Si/SiO$_2$ substrate with the use of the direct technique of amplitude-modulated absorption of sub-THz radiation (AMAR) [26-28]. We show that WSi film on Si/SiO$_2$ substrate is the perfect example of an important class of acoustically soft ultra-thin films on rigid substrates, with a strong internal phonon bottleneck effect. We make the account of this effect to interpret the data and determine the electron-phonon relaxation time on WSi.

In Section II.A we discuss the specific mechanism of phonon exchange between ultra-thin metal films and a thermal bath, introducing a classification of materials/substrate pairs used in SNSPDs and outlining the important differences between acoustically soft and rigid groups. In Section II.B we present a phenomenological model of a non-bolometric effect of internal phonon bottle-neck. Section III provides a description of the experiment. Finally, Section IV contains a comparison between the theory and the experiment and general discussion.

## II. THEORY

### A. Phonon exchange between ultra-thin metal films and thermal bath

The photoresponse of a thin superconducting film, which is a sensing element of any superconducting radiation detectors, depends on the strength of electron-electron (e-e), electron-phonon (e-ph), phonon-electron (ph-e) interactions and on the effectiveness of the energy exchange with a thermal bath. The last one is characterized by phonon escape time, $\tau_{esc}$. Two extreme regimes of the photoresponse are commonly referred to as bolometric and hot electron. The bolometric regime occurs when $\tau_{esc} \gg \tau_{ph-e}$, where $\tau_{ph-e}$ characterizes the reabsorption of phonons by the electron system. In this case electrons and phonons are at quasi-equilibrium, acquiring under radiation nearly the same temperature increments, which are determined by the details of the energy exchange with the thermal bath (substrate). In the hot electron regime when $\tau_{esc} \ll \tau_{ph-e}$, and e-e interaction dominates e-ph interaction, $\tau_{ee} \ll \tau_{e-ph}$, none of the photon energy is wasted for detection. Phonons in the film are easily exchanged with phonons in the substrate thus forming the thermal bath for relaxing hot electrons. The relative strength of e-e versus e-ph interactions depends on the characteristic energy of electrons and phonons. Usually it is nonmonotonic, providing energy flow from electrons to phonons and back in the process of spectral down-conversion of non-equilibrium distributions [15].

The slowing down of energy dissipation from a metal film to a substrate occurs because of a well-known effect of phonon bottleneck due to the reabsorption of phonons by the electron system. This effect is determined by the ratio $\tau_{esc}/\tau_{ph-e}$. In the limit $\tau_{esc}/\tau_{ph-e} \gg 1$ it results in the bolometric photoresponse. The conventional expression for phonon escape time is $\tau_{esc} = 4d/\eta c$, where $d$ is the film thickness, $\eta$ is the phonon transmission coefficient across the film/substrate interface and $c$ is the mean sound velocity in the film. This expression for $\tau_{esc}$ is easily derived by relating the rate of change in number of phonons in the film to the normal component of phonon flux through the escape interface. The implicit assumption in this derivation is that the incident phonons are isotropically distributed at all instances of time.

For ultra-thin films this assumption becomes questionable at least for certain metal/substrate pairs. Indeed, for an acoustically "soft" metal film on a rigid substrate there is a total internal reflection of phonons with incident angles, $\vartheta$, larger than the critical angle, $\vartheta_C$, i.e. $\vartheta \geq \vartheta_C = \sin^{-1}(c/c_{sub})$ where $c_{sub}$ is the mean sound velocity of the substrate. As a result, phonons with $\vartheta \geq \vartheta_C$ are restricted to propagate within the plane-parallel film until they undergo a scattering-mediated conversion into the part of phase space within the critical cone $\vartheta \leq \vartheta_c$. On the other hand, phonons, which are within the critical cone, escape from the film to the substrate on average after $\tau_{esc,C} = 2d/\bar{\eta}_C c$, where $\bar{\eta}_C$ is the characteristic transmission coefficient for $\vartheta \leq \vartheta_C$. The dependence of $\eta_C(\vartheta)$ is weak for most of the range $\vartheta < \vartheta_c$ decreasing steeply to zero on approach to $\vartheta_C$. Close to normal incidence the transmission coefficient is $\eta_C(\vartheta) \approx 4z/[1+z]^2$, where $z$ is the ratio of acoustic impedances of the film and the substrate. For all typical metal/substrate pairs $\bar{\eta}_C > 0.5$. For ultra-thin metal films used for SNSPD taking $d = 4$ nm, and $c = 3 \cdot 10^5$ cm/s, we obtain $\tau_{esc,C} \cong 5.4$ ps for phonons within the critical cone and $\tau_C(\vartheta \geq \vartheta_C) = \infty$ for phonons outside the critical cone. The time $\tau_{esc,C}$ can be associated with the depletion time for phonon occupation numbers within the critical cone. Conventional estimate for $\eta \leq 0.3$ yields an average escape time for all phonons $\tau_{esc} = 4d/\eta c \geq 18$ps.

It is evident from the estimates above that in the extreme case of a soft ultra-thin metal film on a rigid substrate phonons are discriminated into groups of escaping (within the critical cone) and non-escaping modes (outside the critical cone). Correspondingly, non-equilibrium distributions for these groups are totally different. One group is directly coupled and exchanges energy with the substrate, while another is de-coupled from it. Neglecting this discrimination and describing the phonon system as a whole via the introduction of averaged characteristics means disregarding the important details of the energy exchange between the film and the substrate. As we demonstrate below, such a discrimination results in qualitative changes of the energy exchange rates. However, the conventional approach to describing phonons as a unified system is still applicable while "average" escape time, $\tau_{esc}$, is within the limited range $\tau_{esc,C} < \tau_{esc} \ll \infty$. In this case $\tau_{esc}$ can be viewed as a free parameter accounting on average for a quick energy exchange for escaping and the absent exchange for non-escaping phonon groups. In Section IV we will review the implications of using the conventional theory of a phonon bottleneck effect in ultra-thin films beyond its validity limits.

Let us now analyse ultra-thin superconducting films, which are currently used for SNSPDs, grown on conventional dielectric substrates. Table I summarises the known elastic properties of some of the used materials and the substrates.

Table I. Material characteristics

| Material | $T_C$ (K) | $\rho$ (g/cm$^3$) | $c_t$ ($10^5$ cm/s) | $c_l$ ($10^5$ cm/s) | $c$ ($10^5$ cm/s) | $T_D$ (K) | Ref |
|---|---|---|---|---|---|---|---|
| $Nb$ | 4.2 | 8.59 | 2.17 | 5.14 | 2.45 | 275 | [24, 29] |
| $NbN$ | $\leq 16$ | 8.47 | 4.2–5.0 | 7.8–8.4 | 4.7–5.6 | 800–950 | [30] |
| $NbC$ | $\leq 11$ | 7.82 | 5.1 | 8.4 | 5.6 | 740 | [31] |
| $NbSi_2$ | 2.9 | 5.7 | 5.2 | 8.3 | 5.7 | 688 | [32] |
| $TaN$ | $\leq 10.8$ | 14.3 | 3.6 | 6.4 | 4.0 | 519 | [33,34] |
| $W_3Si$ | $\leq 5$ | 15.8 | 3.0 | 5.4 | 3.3 | 391 | [35] |
| $W_5Si_3$ | $\leq 4.5$ | 14.2 | 2.9 | 5.5 | 3.2 | 384 | [36] |
| $WSi_2$ | $\leq 1.9$ | 9.8 | 4.5 | 7.1 | 5.0 | 610 | [37] |
| $Mo_{0.8}Si_{0.2}$ | $\leq 7.5$ | 9.2 | 2.7 | 6.1 | 3.0 | 364 | [38,39] |
| $Mo_{0.8}Ge_{0.2}$ | $\leq 7.4$ | 9.7 | | | | 266 | [40] |
| $Si$ | | 2.3 | 5.3 | 9.0 | 5.9 | 640 | [29] |
| a-$SiO_2$ | | 2.48 | 3.4 | 5.4 | 3.8 | 476 | [29] |
| $Al_2O_3$ | | 3.99 | 6.6 | 10.9 | 7.1 | 1028 | [29] |
| $GaAs$ | | 5.31 | 3.0 | 5.2 | 3.3 | 345 | [29] |

As seen from Table I all metals can be split into two groups. All Nb compounds are characterised by mean sound velocities exceeding $5.0 \cdot 10^5$ cm/s. All other intermetallic compounds: WSi, MoSi, MoGe and elemental Nb form another group of acoustically softer materials with mean sound velocities close to or less than $3.0 \cdot 10^5$ cm/s, TaN being intermediate. Note, that most of the data in Table I are shown for the ordered phases and were calculated using density functional techniques. Only in very few cases the experimental measurements of elastic properties are available, and very rarely for thin films. Substantial acoustic softening was experimentally observed in $Mo_{1-x}Si_x$ at the moment when the alloy becomes amorphous at $x \approx 0.19$ [39]. Composition with $x \approx 0.2$ is optimal for superconducting properties of MoSi. The same is true for amorphous MoGe [40]. For WSi sound velocities in ordered phases $W-W_3Si-W_5Si_3$ change weakly with the increase of Si content and exhibit significant rise from $W_5Si_3$ to Si rich $WSi_2$. We expect that the same behaviour remains true in amorphous material.

The acoustic properties of the shown substrates also cover a wide range. Therefore, many options of metal/substrate matching are available. Nb compounds can be well matched with few substrates, providing a quick energy exchange with thermal bath. Another situation arises for amorphous WSi, MoSi, MoGe grown on fast substrates, like $Al_2O_3$ or Si and to a lesser extent on $SiO_2$. These pairs are acoustically mismatched, affecting energy transport to a thermal bath. Below we analyse WSi on silicon, silicon dioxide and sapphire substrates to illustrate implications of acoustic mismatch on the flow of energy. We will not consider WSi/GaAs, which is the opposite example of a good match. For $WSi/Al_2O_3$, WSi/Si and $W/SiO_2$ we estimate $\vartheta_C = 27°, 34°$ and $60°$ respectively, so that corresponding volumes of the phase space within the critical cones become 0.11, 0.17 and 0.50 of the full phase volume. In all these cases we expect the splitting of phonons into escaping and non-escaping groups. Since this substantially influences the energy flow in ultra-thin films, the depletion of phonon states within the critical cone cannot be neglected. The straightforward condition for this to happen can be written by balancing the out-flow rate of the phonons from the critical cone to in-flow rate from outside of the critical cone. The out-flow rate is $\tau_{esc,C}^{-1}$. The in-flow rate is controlled by processes of the scattering-mediated conversion which connect phonons of the two groups. One is the direct process with the rate $\tau_s^{-1}$ either due to elastic bulk or interface/surface roughness scattering. Another conversion process is sequential, and involves the absorption of one of the non-escaping phonons by an electron with the subsequent emission of another phonon into the critical cone. In both cases the probability of the phonon to arrive in the escaping group is proportional to the fraction of the phase space inside the critical cone.

The formal validity criterion of the conventional model for soft metals on a rigid substrate thus becomes $d > d_{th}$, where the threshold thickness of the film is

$$d_{th} \approx \frac{\bar{\eta} c}{2} \left( \min\{ \tau_{e-ph}^{-1}, \tau_{ph-e}^{-1} \} + \tau_s^{-1} \right)^{-1}.$$

The strong scattering at a diffuse interface or in bulk with the elastic mean free path smaller or of the order of few nm would maintain isotropic phonon distribution and validate the use of the conventional model of a phonon bottle-neck. However, this case is unlikely, because at low temperatures the wavelengths of thermal phonons exceed the characteristic spatial scales of possible structural defects or interface (surface) roughness. More realistic is the opposite limit of relatively mild elastic scattering, $l_s = c\tau_s \gg d$. Taking $\tau_{e-ph} = 66$ ps for WSi at 4.5 K [21] and neglecting elastic scattering we obtain $d_{th}(\tau_s \to \infty) \approx 50$ nm. Weak elastic scattering, $l_s = c\tau_s \gg d$, results in $d \ll l_s$, and $d \ll d_{th} = min\{l_s, d_{th}(\tau_s \to \infty)\}$. Thus, for ultra-thin WSi films the conventional model of a phonon bottleneck is not valid.

Although the scenario of an internal phonon bottleneck was discussed under the assumption of 3D phonons in the film, the same scenario also makes sense for thin films with phonons of reduced dimensionality. In this case we may split phonons into two groups as well – leaking and non-leaking. The first group provides channels for the energy exchange with the thermal bath. The second group is connected with escaping modes via direct or indirect conversion mechanisms. These mechanisms play similar roles to phonons inside and outside of the critical cone in 3D case.

### B. Model of internal phonon bottle-neck

To analyze the energy exchange between an acoustically soft metal film on a rigid substrate we adopt a simple phenomenological model. In this model we describe interactions between the electron system and the two phonon subsystems. Each of them is at quasi-equilibrium and characterized by the respective temperature. Hereafter we will refer to that as three-temperature (3T) model.

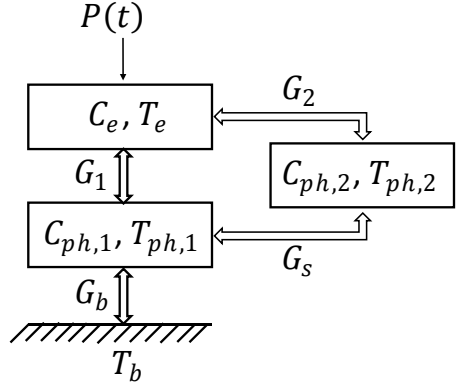

Fig. 1. Energy balance diagram.

Fig. 1 schematically illustrates the energy balance between the thin metal film and the substrate. $P(t)$ is the incident power, $C_e, C_{ph,1}, C_{ph,2}$ are heat capacitances of electrons and phonons of groups 1 and 2 respectively, $G_1$ and $G_2$ are thermal conductances describing e-ph interactions connecting electrons and phonons of corresponding groups, $G_s$ – conversion conductance describing the energy exchange between two groups of phonons, $G_b$ – thermal conductance describing the energy exchange between escaping (group 1) phonons and the thermal bath, and finally $T_e$, $T_{ph,1}$, $T_{ph,2}$ and $T_b$ are the temperatures of corresponding systems. As in the conventional two-temperature (2T) model, such a description is based on the assumption that the phonon system (phonon systems in 3T-model) can be described by their own temperature(s). Although such an assumption cannot be rigorously justified, nonetheless the integral energy balance is usually described fairly well, at least for the 2T-model. We write down the equations of the 3T-model in the form

$$C_e \frac{dT_e}{dt} = -G_1(T_e - T_{ph,1}) - G_2(T_e - T_{ph,2}) + P(t)$$
$$C_{ph,1} \frac{dT_{ph,1}}{dt} = G_1(T_e - T_{ph,1}) - G_b(T_{ph,1} - T_b) + G_s(T_{ph,2} - T_{ph,1}) \quad (1)$$
$$C_{ph,2} \frac{dT_{ph,2}}{dt} = -G_2(T_e - T_{ph,2}) - G_s(T_{ph,2} - T_{ph,1}).$$

Introducing the total thermal conductance for e-ph interaction $G = G_1 + G_2$ and assuming isotropic scattering, we obtain $G_1 = G \frac{C_{ph,1}}{C_{ph}}$ and $G_2 = G \frac{C_{ph,2}}{C_{ph}}$, where $C_{ph} = C_{ph,1} + C_{ph,2}$ is the total phonon heat capacity. Note that this is a simplification and such relations do not hold for 2D phonons, where coupling between electrons and leaking or non-leaking 2D-phonons in general is not the same.

Balance equations (1) were written for small deviations from equilibrium: $P(t) = P_0 + \delta P(t)$, $T_e(t) = T_{e,0} + \delta T_e(t)$, $T_{ph,1}(t) = T_{ph,10} + \delta T_{ph,1}(t)$, $T_{ph,2}(t) = T_{ph,20} + \delta T_{ph,2}(t)$, $T_{e,0} \approx T_{ph,10} \approx T_{ph,20}$ and $|\delta T_e(t)| \to 0$, $|\delta T_{ph,1}(t)| \to 0$, $|\delta T_{ph,2}(t)| \to 0$. We consider the linear response to an oscillating input power, $P(t) = P_0 \cos \omega t$, thus taking $\delta T_e(t) = \delta T_e(\omega) e^{-i\omega t}$. The characteristic relaxation times in terms of heat conductances and capacitances in Eqns. (1) are introduced according to

$$\frac{1}{\tau_{e-ph}} = \frac{G}{C_e}, \quad \frac{1}{\tau_s} = \frac{G_s}{C_{ph,2}}, \quad \frac{1}{\tau_{esc}} = \frac{G_b}{C_{ph,1}}.$$

The solution of Eqns. (1) depends on five independent parameters: electron-phonon relaxation time, $\tau_{e-ph}$, conversion time of non-escaping (group 1) into escaping (group 2) phonons, $\tau_s$, escape time of phonons of group 1, $\tau_{esc}$, the ratio of electron to phonon heat capacities, $C_e/C_{ph}$, and the ratio of heat capacity of escaping phonons to total phonon heat capacity, $C_{ph,1}/C_{ph}$. The conventional 2T-model contains only three independent parameters, namely, $\tau_{e-ph}$, $\tau_{esc}$ and $C_e/C_{ph}$.

In terms of the 3T-model it is now useful to consider several limiting cases for variety of "soft" superconducting films on rigid substrates listed in Table I. Moreover, this will reflect how the model is sensitive to various parameters. According to our experimental approach, the analysis should be done by considering spectrum of the output signal which is the measure of $T_e$ change. As we will see different input parameters affect both the shape of the simulating curve and position of the roll-off, and also other points of interest, for example the predicted inflection points.

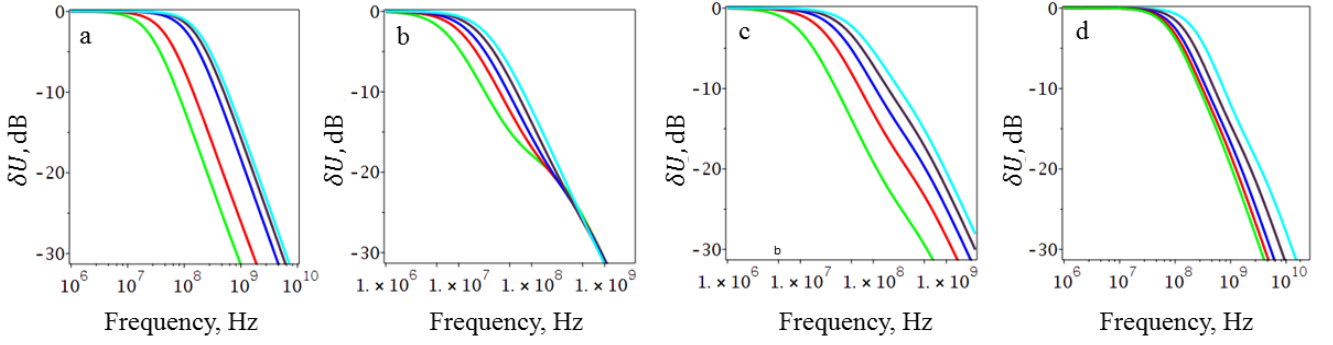

Fig.2. (Color online) Simulation of the spectrum of the output voltage signal with 3T-model as a function of a) conversion time, b) ratio of electron to phonon heat capacity, c) ratio of escaping phonon to total phonon heat capacity, d) electron-phonon relaxation time.

Fig. 2a shows the curves simulated for varied conversion time $\tau_{e-ph}/\tau_s = 0, 1, 10, 100$ from left to right, the other parameters are: $C_e/C_{ph} = 5.0$, $\tau_{e-ph} = 800$ ps, $C_{ph,1}/C_{ph} = 0.15$. The outmost right curve shows the predictions of the 2T-model. Fig. 2b demonstrates the dependence on the ratio of electron to phonon heat capacities $C_e/C_{ph} = 0.3, 0.6, 1.0, 2.0, 5.0$ from left to right, the other parameters are: $\tau_{e-ph} = 800$ ps, $\tau_{e-ph}/\tau_s = 0$, $C_{ph,1}/C_{ph} = 0.15$. Fig. 2c illustrates sensitivity to the phase volume of escaping phonons with $C_{ph,1}/C_{ph} = 0.1, 0.2, 0.3, 0.4, 0.5$ from left to right, and the other parameters are: $\tau_{e-ph} = 800$ ps, $C_e/C_{ph} = 1.0$, $\tau_{e-ph}/\tau_s = 0$. Finally, Fig. 2d shows curves for varied electron-phonon relaxation time $\tau_{e-ph} = 100, 200, 400, 600$ and $800$ ps from right to left, and the other parameters are: $C_e/C_{ph} = 1.0$, $C_{ph,1}/C_{ph} = 0.2$ and $\tau_s = 1$ ns. Not shown in any of the figures is the value of escape time, which was chosen $\tau_{esc} = 3$ ps. For the geometry of our film the traverse time for escaping phonons is $d/c \approx 1.0$ ps. Therefore, $\tau_{esc} = 3$ ps is a sensible value for phonons of group 1. As long as $\tau_{esc} \leq 10$ ps the results of simulations are not sensitive to its true value. Thus, the 3T-model operates with four rather than five independent parameters. The drop-out of $\tau_{esc}$ has clear meaning: at chosen parameters of scattering rates, bottle-necking in energy relaxation can occur between the electron and any of the phonon subsystems or between the two phonon subsystems, but not between the subsystem of escaping phonons and the substrate.

Taking in Eqns. (1) $C_{ph,1}/C_{ph} = 1$, $G_1 = G$, $G_2 = 0$ or the limit of infinitely fast conversion rate, $\tau_s \to 0$, one will obtain the conventional 2T-model. Performing simulations for the 2T-model for the wide range of parameters we only observe the shift of the roll-off frequency and no change of their shapes. Note that for the 2T-model, as discussed in Section II.A, we must allow $\tau_{esc}$, to change over substantially wider range. In fact, using similar parameters as in Fig. 2 and allowing $\tau_{esc}$ for the conventional model to vary between 10 and 100 ps we obtain the sets of curves shown in Fig. 3. It is worth noting that for WSi with $d = 4$ nm and $c = 3.3 \cdot 10^5$ cm/s $\tau_{esc} = 100$ ps would correspond to unlikely $\eta = 0.05$, i.e. bad bonding of the film to the substrate.

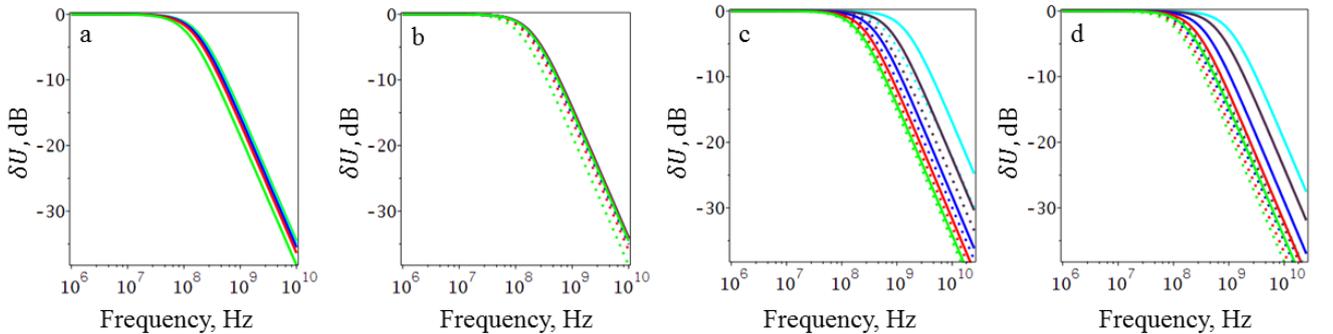

Fig.3. (Color online) Simulation of the spectrum of the output voltage signal with the conventional 2T-model for different values of phonon escape time a) $\tau_{esc} = 10, 20, 30, 50, 100$ ps from right to left, $C_e/C_{ph} = 5.0$, $\tau_{e-ph} = 800$ ps, b) solid curves − $\tau_{esc} = 10$ ps, dots − $\tau_{esc} = 100$ ps, $C_e/C_{ph} = 0.3, 0.6, 1.0, 2.0, 5.0$ from right to left $\tau_{e-ph} = 800$ ps, c) solid curves − $\tau_{esc} = 10$ ps, dots − $\tau_{esc} = 100$ ps, $\tau_{e-ph} = 100, 200, 400, 600, 800$ ps from right to left, $C_e/C_{ph} = 1.0$ , d) same as c) but with $C_e/C_{ph} = 5.0$.

Both the proposed 3T-model and the conventional 2T-model predict qualitatively similar shifts of the roll-off frequencies with the change in the material parameters. However, we note the possibility of substantial variations of shapes in Fig. 2 (3T-model), while shapes in Fig. 3 (2T-model) remain unchanged. Since variations of curvature seen in Fig. 2 occur at higher frequencies, these features may be experimentally missed if the frequency range beyond the roll-off frequency is narrow. Moreover the position of the roll-off frequency predicted by the 2T- and 3T-models substantially differs for all values of the expected parameters. Thus, fitting the experimental data over a range of frequencies where slopes remain unchanged (immediately above the roll-off frequency for the 3T-model) is possible also with the 2T-model even outside its validity range. However, this will involve the use of parameters which are difficult to justify. In Section IV when comparing our results to the experiment we will show that this is exactly the case.

## III. EXPERIMENT AND RESULTS

### A. Sample preparation

A superconducting WSi film was grown on a thermally oxidized silicon substrate at room temperature using DC magnetron sputtering in an argon (Ar) atmosphere. The substrate was placed onto a rotatable holder and the holder was alternately positioned over pure tungsten (W) and pure silicon (Si) targets. By sequential deposition of ultra-thin W and Si layers over seven revolutions of the holder, the resulting WSi film (atomic ratio W:Si=3:1 calibrated by a sputtering rate of individual components) of thickness $d = 3.4$ nm was obtained. To prevent oxidation the WSi film was capped with a 4 nm thick Si layer. Transmission electron microscopy analysis of the deposited film showed that it is amorphous. The film was patterned into a microstrip of a width $w = 1$ μm and length $L = 10$ μm using electron-beam lithography and placed between V-Cu contact pads. The critical temperature of the patterned film was 3.4 K, and the width of resistive transition was 0.25 K.

### B. Experimental technique

In this work in order to study the electron-phonon relaxation we use the AMAR technique. The sample is kept at the superconducting transition temperature (where even a slight shift of temperature significantly changes the resistance), $T_C$, biased with low DC current, and exposed to an amplitude-modulated radiation of sub-THz range. The absorption of incident modulated radiation causes an increase of the electron temperature, $T_e$, and consequently results in an increase of the sample resistance, producing a change of the voltage signal. The amplitude of the voltage signal, $\delta U$, exhibits frequency-dependent roll-off (inset of spectrum analyzer in Fig. 4), which is used as a measure of the sample response time, $\tau_R$. Experimentally obtained $\tau_R$ coincides with the electron-phonon relaxation time if phonons of the film act as a thermal bath. If this condition is not satisfied, the response time is affected by phonon reabsorption. The determination of electron-phonon relaxation time requires special analysis. It is more complicated, and other parameters entering the model must be assessed independently. Usually the energy relaxation processes are studied under quasi-equilibrium conditions, i.e. for low excitation RF and DC power, when variations of the effective temperatures of electron and phonon subsystems are much smaller than the bath temperature. In practice the AMAR technique allows studying both the low heating regime, where the electron-phonon system is under quasi-equilibrium conditions, and the intense heating regime, where $T_e$ is significantly varied. These two regimes differ only in the amount of the incident RF power, hereafter referred to as *low and high RF power regimes*.

An experimental setup is schematically depicted in Fig. 4. The sample was mounted on a waveguide flange and placed into liquid helium cryogenic insert which operates in a temperature range down to 1.8 K. The temperature was registered by a carbon thermometer in close vicinity of the sample. Radiation sources used in the experiment are the two backward wave oscillators (BWOs), operating at close frequencies, $f_1$ and $f_2$, in the range of 120-145 GHz. RF signals were coupled by a 50/50 beam splitter, and passed through waveguide (with a polarization controller) to the sample. Interference between RF signals causes the beating of the optical power with intermediate frequency (IF), $f_{IF} = f_2 - f_1$. A shift of the IF from 5 MHz up to tens of GHz was realized by sweeping the frequency of the local oscillator (LO) BWO, while the frequency of the signal BWO was constant. Since the LO power depends on frequency, using the attenuator we adjusted the LO power every time the frequency was changed. The sample response was passed through the bias-tee with a separated RF path and DC bias, amplified at room temperature, and, together with the IF, measured by the spectrum analyzer. The readout path was calibrated during the experiment.

The typical spectrum of the voltage response is shown in the spectrum analyzer inset in Fig. 4. It exhibits the expected plateau for low frequencies and high frequency roll-off. To obtain the response time, we fitted the experimentally measured spectrum of the voltage response according to the expression

$$\delta U(f) = \delta U(0)\left(1 + \frac{f^2}{f_{3dB}^2}\right)^{-1/2}, \quad (2)$$

where $\delta U(0)$ and $f_{3dB} = 1/(2\pi\tau_R)$ are fitting parameters, $f_{3dB}$ is 3dB roll-off frequency of the response spectrum.

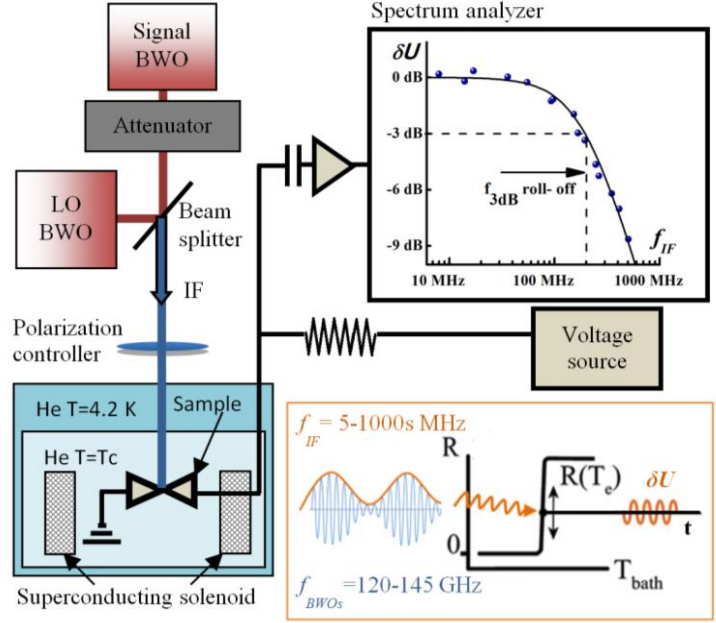

Fig.4. (Color online) The experimental setup for measurements of energy relaxation. The sample which is kept in the middle of the superconducting transition absorbs amplitude-modulated radiation at an intermediate frequency from the two BWOs mixing on the sample. The change of electron temperature caused by the absorbed radiation induces the change of sample resistance that consequently generates an output signal, $\delta U$, which is measured by the spectrum analyzer. The frequency dependence of $\delta U$ is used to define the roll-off frequency at -3 dB level and determine the response time $\tau_R$.

In the *low RF power regime* the electron-phonon system is under quasi-equilibrium conditions. To check that this is the case, we studied $\tau_R$ as a function of DC and RF power, Fig. 5a. We maintained the sample in the middle of the superconducting transition by irradiating it with amplitude-modulated radiation of low power and applying a low DC current. The response time from the measured $\delta U(f)$ is plotted as a function of DC voltage in Fig. 5a (red solid symbols). Within the voltage range, where the change of $\tau_R$ does not exceed 10%, we chose the bias, at which the output signal was maximal. Therefore, for further measurements, the voltage bias was limited to the linear regime. Likewise, we measured $\tau_R$ as a function of RF power (Fig.5a blue open symbols). Within the range of RF powers, where the change of $\tau_R$ did not exceed 10%, we similarly chose the RF power, at which the output signal was maximal. Thus the RF power was also restricted to the linear regime. For the found operational parameters, the DC and RF heating are minimal and assumed not affect the energy relaxation rate of the electron system. Current-voltage characteristics of the microstrip are plotted in Fig. 5b, where the solid blue curve was obtained in a superconducting state at 1.8 K and the dashed black curve (the operational IV curve) was obtained at $T_C$ irradiating the microstrip by low RF power.

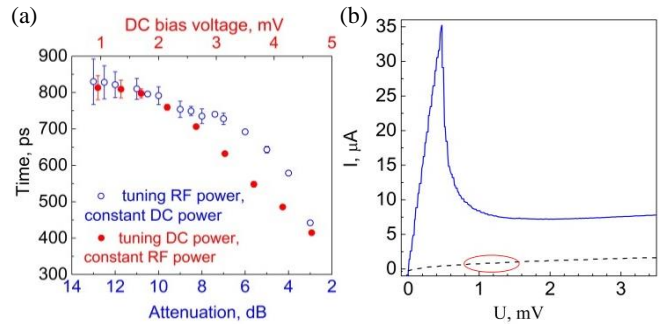

Fig.5. (Color online) Establishing the low RF power regime: (a) $\tau_R$ as a function of DC voltage at fixed value of RF power (red solid symbols) and as a function of incident RF power at fixed value of DC voltage (blue open symbols). The range where $\tau_R$ does not depend on DC bias voltage and the attenuated RF corresponds to negligible DC and RF heating. The error bars give the uncertainty of the roll-off frequency. (b) I-V characteristic. Dashed and solid curves refer to a normal-superconducting transition and a superconducting state at T=1.8 K respectively. The selected area is the operational range of voltages.

The experimentally measured relaxation time is an intrinsic property of the relaxation processes in the film, and is not affected by the electro-thermal feedback (ETF), in contrast to the transition edge sensors (TES) [41]. The response time corrected due to ETF is given by

$$\tau_R' = \frac{\tau_R}{1+\mathcal{L}_0/(1+\alpha_I)}. \qquad (3)$$

The correction is determined mainly by a loop gain factor $\mathcal{L}_0 = P_0 \alpha_T/GT_0$, where characteristic power $P_0$ is expressed in terms of the bias current $I_0$, resistance of the sample at the operating point, $R_0$, and the load resistance, $R_L$, as $P_0 = I_0^2(R_0 - R_L)\kappa$ [42] with $\kappa = (R_0 - R_L)/(R_0 + R_L)$; $G$ is the thermal conductance, $\alpha_T = (T/R_0) \cdot (dR/dT)$ and $\alpha_I = (I_0/R_0) \cdot (dR/dI)$. The load resistance was 50 Ω. From IV curve (the dashed curve on Fig. 5b), $R_0 \approx 1$ kΩ, while from the $R(T)$ curve (see Fig. 8) we obtain the rough estimate $\alpha_T \approx 10$. Estimating thermal conductance as $G = C_e/\tau_R$, we obtain the negligibly small loop gain factor, $\mathcal{L}_0 \approx 0.05$. The account for the non-zero logarithmic derivative of resistance versus current, $\alpha_I$, can only decrease the correction. The use of smaller $\tau_{e-ph}$ instead of $\tau_R$ in the estimate of $G$ increases $G$ and decreases loop gain factor. Thus, we neglect the effect of ETF.

Once the correct operational parameters have been found, we further studied the temperature dependence of $\tau_R$ in the low RF power regime. Since the AMAR technique is applied only at $T_C$, we shifted the transition temperature by applying a magnetic field perpendicular to the sample surface. To create the magnetic field we used superconducting solenoid mounted together with cryogenic insert into He-Dewar (Fig. 4). As illustrated in Fig. 6, the applied magnetic field shifts the signal $\delta U$ to lower frequencies, resulting in increase of $\tau_R$. The response time is plotted as a function of temperature in the inset of Fig. 6. Since $\tau_R$ is expected to scale as $\tau_R \propto T^{-n}$, we obtain $n = 2.87 \pm 0.18$, which can be identified with the integer value $n = 3$. At zero magnetic field $\tau_R = 800$ ps.

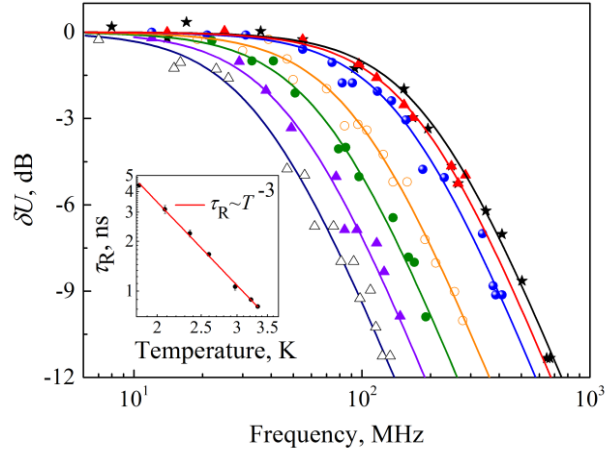

Fig.6. (Color online) $\delta U$ vs frequency in low RF power regime for different magnetic fields: from left to right 3.25, 3, 2.38, 2.1, 1.4, 1.1, 0.72 T which correspond to following transition temperatures: 1.8, 2, 2.4, 2.6, 3, 3.2, 3.3 K accordingly. Solid curves are fits to data using Eq. (2). Inset: Temperature dependence of the response time $\tau_R$. The solid line is fit according to $\tau_R \propto T^{-n}$, where $n = 3$.

Since the response amplitude is proportional to the absorbed RF power, in low RF power regime the amplitude of the response is small. This results in strong limitations of its validity. To expand the spectral range we performed AMAR measurements in *the high RF power regime*. We increased the power of both the local and the signal BWOs as much as possible and kept it constant during the measurement. High RF power caused an increase of the electron temperature compared to the bath temperature. To take this into account, we controlled $T_e$ by measuring roll-off time, $\tau_R$, and comparing it to that measured in the low RF power regime (inset in Fig. 6), i.e. using $\tau_R$ as the thermometric value. Such an approach allowed us to measure the voltage response down to -25 dB expanding the spectral range by an order of magnitude. The experimental results and simulations based on the 3T-model for temperatures 2.65 K and 3.4 K are shown in Fig. 7. We leave a detailed description of the comparison of the theory and the experiment for the next section.

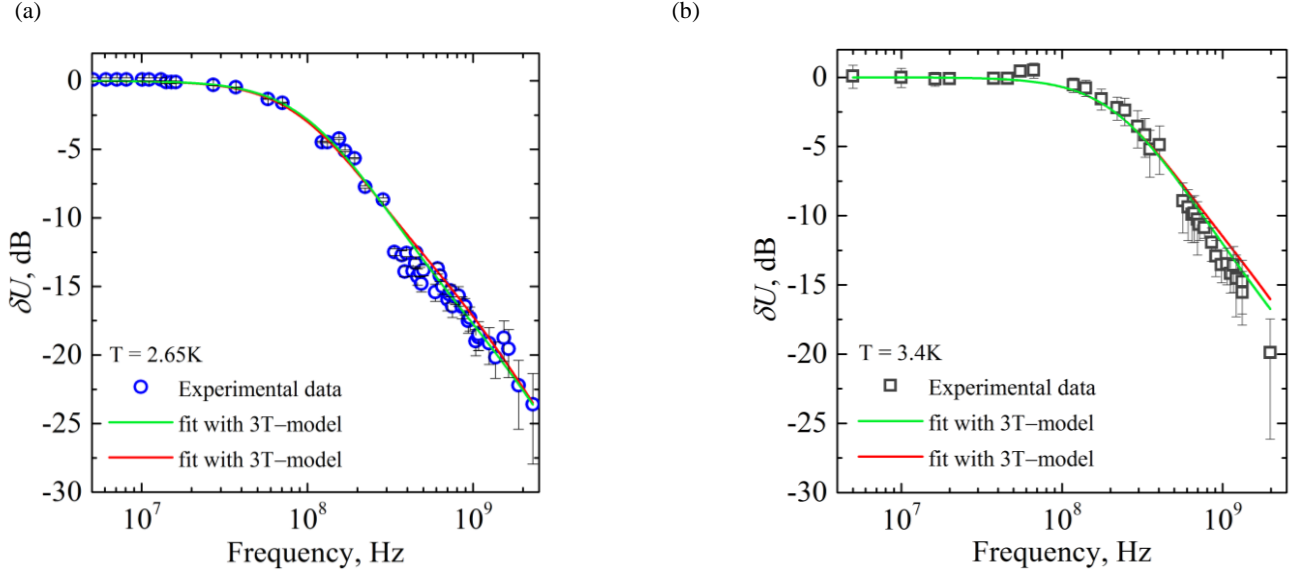

Fig. 7. (Color online) $\delta U$ vs frequency in high RF power regime at (a) 2.65 K and (b) 3.4 K. Open symbols - experimental data, solid curves - fits according to 3T-model. Temperature-independent fitting parameters - red curve $C_e/C_{ph} = 0.86$, $C_{ph,1}/C_{ph} = 0.15$, $\tau_{esc} = 2$ ps - green curve $C_e/C_{ph} = 1.1$, $C_{ph,1}/C_{ph} = 0.165$, $\tau_{esc} = 2$ ps. Values of temperature dependent parameters $\tau_{e-ph} \sim T^{-3}$ and $\tau_s \sim T^{-4}$ (Rayleigh law) at 2.65 K - red curve, $\tau_{e-ph} = 400$ ps and $\tau_s = 1.1$ ns - green curve, $\tau_{e-ph} = 290$ ps and $\tau_s = 1.3$ ns. Error bars denote statistical uncertainties from noise level.

To compare the obtained results with the results of other groups and theoretical predictions, we carried out measurements of the electron diffusion constant, $D$, and electron density of states at Fermi level, $\nu(0)$. $D$ was obtained from the dependence of the second critical magnetic field, $H_{C2}$, on temperature close to $T_c$. The introduction of a magnetic field results in a reduction of the critical temperature and a widening of the transition, shifting the $R(T)$ curves to lower temperatures (Fig.8). The sample was biased with a small alternating current, which did not cause any noticeable heating of the electron subsystem. The diffusion constant was calculated according to formula [43]:

$$D = 1.097 \left(-\frac{dH_{c2}}{dT}\right)^{-1}\bigg|_{T=T_C}, \qquad (4)$$

where $dH_{c2}/dT$ was calculated from the slope of the linear fit to the experimental data, see inset in Fig.8. The calculated $D$ for the WSi thin film using eq.(4) is 0.58 cm$^2$/s. The electron density of states at the Fermi level was estimated using the experimentally determined values of the electron diffusion constant and the resistivity according to Einstein relation $\nu(0) = 1/(e^2\rho D)$, and found to be $5.3\times10^{22}$ eV$^{-1}$ cm$^{-3}$.

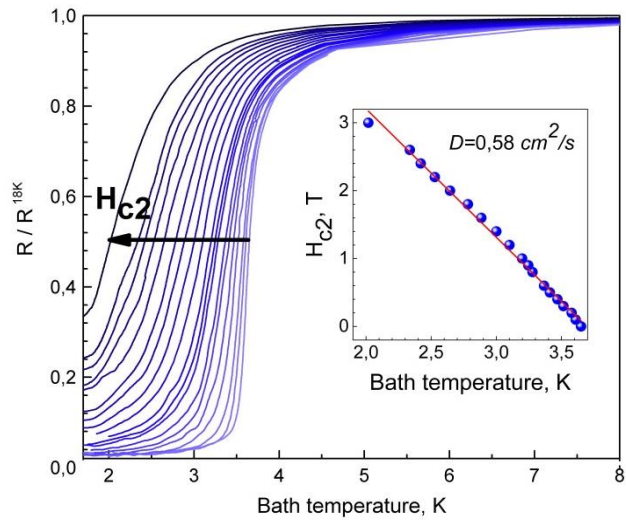

Fig.8. (Color online) Temperature dependence of sample resistance for different magnetic fields. The inset shows the second critical magnetic field vs temperature.

The measured parameters of WSi are summarized in Table II. $R_\square^{300K}$ is the sheet resistance at room temperature, the resistivity was determined from the sheet resistance and film thickness, $d$, as $\rho^{300K} = dR_\square^{300K}$. The critical current density, $j_C$, was calculated using data of the critical current of 35.2 µA at temperature 1.8K (solid blue curve Fig. 5b).

TABLE II. Parameters of the WSi film.

| $d$, (nm) | $L$, (µm) | $w$, (µm) | $T_C$, (K) | $\delta T_C$, (K) | $R_\square^{300K}$, ($\Omega$) | $\rho^{300K}$, ($\Omega$ cm) | $j_C$ ($T$=1.8K), (A / cm$^2$) | $D$, (cm$^2$ / s) | $\tau_R$ at $T_C$ (ps) | $\nu(0)$, (eV$^{-1}$ cm$^{-3}$) |
|---|---|---|---|---|---|---|---|---|---|---|
| 3.4 | 10 | 1.0 | 3.4 | 0.25 | 595 | 2.02×10$^{-4}$ | 1.04×10$^6$ | 0.58 | 800 | 5.3×10$^{22}$ |

## IV. DISCUSSION

### A. Fitting procedure

The experimental results in Fig. 7 substantially deviate from the predictions of the conventional 2T-model. At 2.65 K at a frequency close to $f \sim 4 \div 5 \cdot 10^8$ Hz we see a feature reminiscent of the inflection point. Such a feature is smeared at 3.4 K. Despite a significant scatter of points, the observed feature cannot be attributed to a systematic error of measurement. Firstly, it is not pinned to a certain frequency and changes both its position and shape with temperature, hence it is not due to a calibration of the RF path. Secondly, it cannot occur due to a nonlinearity of the response, in which case it should be more pronounced at high amplitudes of the output signal that is contrary to the experiment. In Fig. 9 we plotted the same experimental data as in Fig. 7 together with the best fit curves in a semi-log scale in order to emphasize the difference between fits of the proposed 3T- (red and green solid curves) and the conventional 2T- models (blue and purple dashed curves). The fitting in Figs. 7 and 9 was obtained by fixing $\tau_{e-ph}$ and $\tau_s$ at $T$=2.65 K and allowing these parameters to scale with the temperature as $\tau_{e-ph} \sim T^{-3}$ and $\tau_s \sim T^{-4}$ (Rayleigh law), but keeping $C_e/C_{ph}$ constant. The ratio $C_{ph,1}/C_{ph}$ was expected to be temperature independent. We showed the two best quality fits, which we were able to achieve. It is important to emphasize that increasing the ratio $C_e/C_{ph}$ above or decreasing it below unity results in poorer quality fits. A better quality fit at 3.4 K can be achieved allowing the ratio $C_e/C_{ph}$ to depend on temperature. Estimating $C_e$ within the free-electron model as $C_e = (\pi^2/3)k_B^2\nu(0)T$ and $C_{ph}$ within the Debye model as $C_{ph} = 12/5\pi^4 N_0 k_B^2 (T/T_D)^3$, where $N_0$ is the number of ions per unit volume, we obtain $C_e/C_{ph} = 5.65$ at $T = T_C$. The smaller best fit value of $C_e/C_{ph} = 0.86 \div 1.1$ indicates a significant deviation of lattice heat capacity at low temperature in WSi from the Debye model. The ratio $C_e/C_{ph}$ is one of the main parameters describing the energy exchange between interacting electrons and phonons. Evaluating this ratio experimentally is important for an independent estimate of the split of the energy between thermalizing electron and phonon distributions following photon absorption in SNSPD, which affects the detection mechanism (parameter $\gamma$ in ref.[18]).

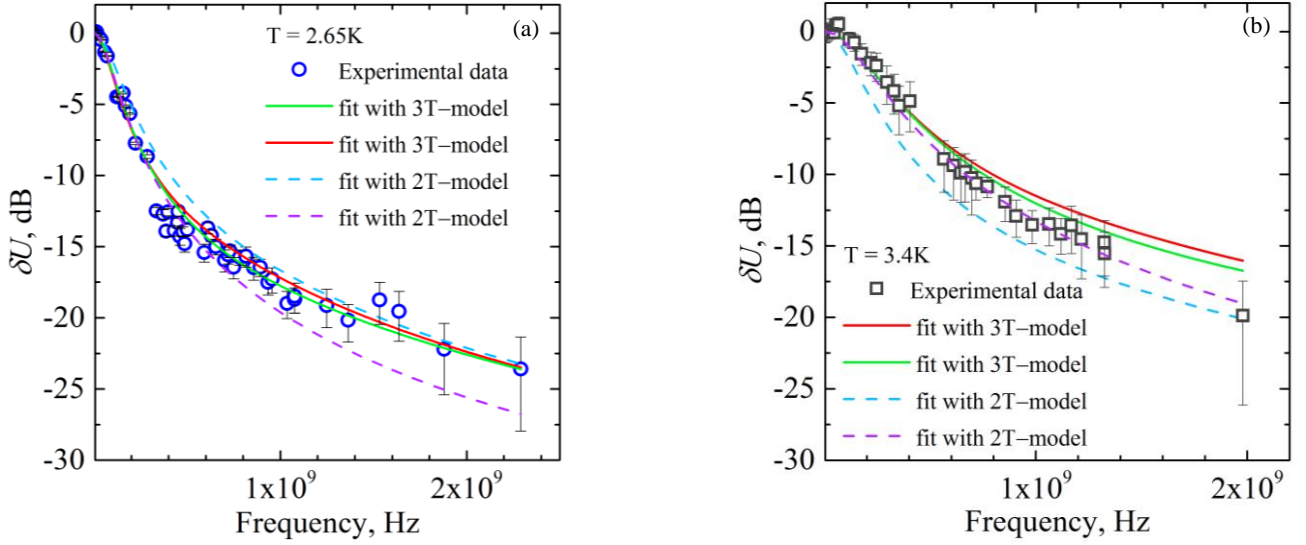

Fig. 9. (Color online) Same as in Fig. 7 but plotted in semi-log scale. Open symbols - experimental data, solid curves - fits according to 3T-model, dashed curves – fits according to 2T-model at (a) 2.65 K, (b) 3.4 K. (a) 3T-model - green solid curve: $\tau_{e-ph}$=290 ps, $\tau_s$=1.3 ns, $\tau_{esc}$=2 ps (group 1 phonons), $C_e/C_{ph}$=1.1, $C_{ph,1}/C_{ph}$=0.165, red solid curve: $\tau_{e-ph}$=400 ps, $\tau_s$=1.1 ns, $\tau_{esc}$=2 ps, $C_e/C_{ph}$=0.86, $C_{ph,1}/C_{ph}$=0.15. 2T-model – blue dashed curve: $\tau_{e-ph}$=500 ps, $\tau_s \to 0$ ($C_{ph,1}/C_{ph}$=1), $\tau_{esc}$=400 ps (average escape time for all phonons), $C_e/C_{ph}$=1.2; purple dashed curve $\tau_{e-ph}$=1500 ps, $\tau_s \to 0$ ($C_{ph,1}/C_{ph}$=1), $\tau_{esc}$=2 ps (average escape time for all phonons), $C_e/C_{ph}$=1. Values of temperature dependent parameters are shown for 2.65 K and scaled as $\tau_{e-ph} \sim T^{-3}$ and $\tau_s \sim T^{-4}$ (Rayleigh law).

## B. Failure of the 2T-model

We now discuss the failure of the 2T-model in more detail. The change of curvature (change of frequency dependence) is clearly seen in the data. It is satisfactorily reproduced by 3T-model simulations in Fig. 9a. In the asymptotic limit (taking either $C_{ph,1}/C_{ph} = 1$, $G_1 = G$, $G_2 = 0$ or considering infinitely fast conversion, $\tau_s \to 0$, as described in Section II.B), we obtain the best fit corresponding to the 2T-model. It is essential that the best fit parameters for the 2T-model involve similar ratio $C_e/C_{ph}$ as the 3T-model. Taking asymptotic limit as described above, i.e. using the equations of the 2T-model, we attempted to achieve as close fits to experimental points as possible allowing the remaining two out of three free parameters of the 2T-model to vary between their extreme values. In this way we obtained the two "best" fits with different sets parameters.

One fit (purple dashed curve Fig. 9) predicts $\tau_{e-ph}$ which is much larger than obtained from other experiments [19, 20, 24]. Moreover, it requires an unrealistically short escape time, $\tau_{esc} \sim d/c \approx 1 - 2$ ps, comparable to time of phonons traversing the film. Such a short escape times are consistent with the 3T-model describing escaping phonons of the group 1, but are in contradiction with the expected "average" $\tau_{esc} = 4d/\eta c \geq 18$ ps for the 2T-model, describing all phonons as explained in Section II. We used the value $\tau_{esc}$=2 ps for group 1 phonons of the 3T-model, the same as for all phonons of the best fit 2T-model (purple dashed curve) for the only reason. The 2T-model is an asymptotic limit of 3T-model. Therefore, the best fit 2T-model (purple curve) is the asymptotic limit ($\tau_s \to 0$ or $C_{ph,1}/C_{ph}$ =1) of the 3T-model with $\tau_{esc}$=2 ps. Another "best" fit by the 2T-model (blue dashed curve) predicts $\tau_{e-ph}$ at 3.4K which is in agreement with independent experiment [19, 20, 24]. However, it predicts wrong temperature dependence of $\tau_{e-ph}$ and requires unexpectedly large escape time $\tau_{esc} = 400$ ps.

To highlight the differences between the 3T- and the 2T- model we analyze fits further, characterizing the goodness of fits with the use of $\chi^2$ statistics. For $T = 2.65$ K the 2T-model fit with adequate value of $\tau_{e-ph}$ (blue dashed curve Fig. 9a) is rejected by $\chi^2$ statistics, with $\chi^2 = 110$ which exceeds the critical value = 72.15 for probability 0.95. 2T-model (purple dashed curve) passes the test with $\chi^2 = 20.9$, $p$-value = 1. For 3T-modeling at 2.65 K we obtained $\chi^2 = 18.35$, $p$-value = 1 (green solid curve) and $\chi^2 = 20.1$, $p$-value = 1 (red solid curve), so that 3T-model passes the test. For $T = 3.4$ K (Fig. 9b) both 2T-models pass the test with $\chi^2 = 4.4$ and $p$-value = 1 (blue dashed curve) and $\chi^2 < 1$, $p$-value = 1 (purple dashed curve), critical value = 47.39. Both 3T-models at 3.4K passes the test with similar statistics, with $\chi^2 < 1$, and $p$-values = 1 (green and red solid curves).

In spite of the closeness of fit of the 2T-model (purple dashed curve Fig. 9b) to the experimental data at 3.4 K it must be rejected. Firstly, it requires unrealistically short phonon escape time. Secondly, it predicts the magnitude of $\tau_{e-ph}$, which is not supported by independent experiments [19, 20, 21], its magnitude is wrong by a large factor ~4-5.

For the second set of the 2T-model best fit curves (blue dashed curves Fig. 9) the magnitude of $\tau_{e-ph}$ at 3.4 K is consistent with [19, 20, 21]. However, this set is rejected by $\chi^2$ statistics for 2.65 K. It also results in a considerable deviation from data in Fig. 9b that can be improved only by assuming the temperature dependence of $\tau_{e-ph}$ to be different to the observed $T^3$ law for any feasible temperature dependence of $C_e/C_{ph}$. Indeed, at a low temperature we have $C_e \sim T$, while $C_{ph} \sim T^3$ or $C_{ph} \sim T^2$ for 3D or 2D phonons respectively, or $C_{ph} \sim T$ (to be consistent with constant $C_e/C_{ph} = 1$ for the range of interest). This results in $\tau_{e-ph} = \tau_R - (C_e/C_{ph})\tau_{esc} \sim a(T/T_C)^{-3} - b(T/T_C)^{-s}$, where a and b are numerical factors of the order of unity and $s = 2, 1$ and $0$ for 3D, 2D phonons or constant $C_e/C_{ph} = 1$ respectively. For $\tau_{e-ph}$ to be substantially smaller than $\tau_R$, the terms $a(T/T_C)^{-3}$ and $b(T/T_C)^{-s}$ must be close in magnitude. Therefore, the temperature dependence of $\tau_{e-ph}$ (assuming $\tau_{esc}$ to be independent of temperature) must significantly differ from $T^{-3}$ law.

As follows from fitting data the most likely value of electron-phonon relaxation time at 2.65 K lies within the range 290 ps $\leq \tau_{e-ph} \leq$ 400 ps. The $\tau_{e-ph} \sim T^{-3}$ law was inferred from magnetoresistance measurements [21] in WSi in the temperature range up to ~20 K. This scaling behavior is consistent with our fitting procedure based on the 3T-model. Using the $T^{-3}$ scaling law to make a comparison to other experiments we estimate 60 ps $\leq \tau_{e-ph} \leq$ 81 ps at 4.5 K, which is consistent with 66 ps reported in [21]. Comparing with another experiment [19, 20] at $T_C$ = 3.7 K we obtain 108 ps $\leq \tau_{e-ph} \leq$ 146 ps. This corresponds to 4.5 ns $\leq \tau_0 \leq$ 6.2 ns and $\tau_0 = \frac{360\zeta(5)}{\pi^2}\tau_{e-ph}(T_C) \approx 37.82\tau_{e-ph}(T_C)$ based on Allen's expression for the rate of energy exchange at low temperatures [44]. The latter expression is straightforward to derive from Eq.(19) of Allen's work introducing $\frac{dE_e}{dt} = C_e \frac{dT_e}{dt} = \gamma T_e \frac{dT_e}{dt} \approx -\gamma T_e \frac{\delta T_e}{\tau_{e-ph}(T_e)}$ and using Kaplan's relation for $\tau_0$ [24]. In a recent paper [19] the value of $\tau_0$ was revised to $\tau_0 \approx 5$ ns after accounting for the difference between switching and critical de-pairing currents bringing it in line with the current work. In [15, 18, 19-20] the kinetics of interacting quasiparticles and phonons was described within the simplified model neglecting the effect of disorder on electron-phonon interaction. Correspondingly for comparison and check of consistency of the results obtained by different techniques, the above estimate of $\tau_0$ was made within the same model.

### C. The temperature dependence of electron-phonon relaxation time

Our analysis based on the energy balance Eqns. (1) does not depend on the model for electron-phonon interaction, the role of disorder, phonon quantization in thin films or the exact structure of the modes supported by thin metal films on a dielectric substrate with or without extra coating. It is also valid for ultra-thin films at low temperatures where the wavelength of thermal phonons may not be small when compared to film thickness. The actual phonon mode structure, however, plays a significant role in electron-phonon interactions and energy relaxation. We also expect that the bottleneck effect for the real spectrum of phonon modes is different from that for the bulk phonons. All of these factors are taken into account within the phenomenological analysis that we used determining the heat capacitances and conductances in Eqs. (1).

Finally, we comment on the observed $\tau_{e-ph} \sim T^{-3}$ temperature dependence of the electron-phonon relaxation time. It is well known that for low frequency phonons satisfying the condition $q_T l \ll 1$, where $q_T = T/\hbar c$ is the thermal phonon wave vector and $l$ is the electron mean free path, the disorder strongly modifies the electron-phonon interaction and $\tau_{e-ph} \sim T^{-4}$ [45-46] assuming bulk (3D) phonons. Quantitatively, for most of the materials from the Table I the ratios of $\tau_{e-ph}$ in the ordered and disordered films at $T \sim T_C$ are of the order of a few tenths - unity (0.37 for $W_3Si$, 0.5-0.9 for NbN with parameters from Table I). At low temperatures scaling law $\tau_{e-ph} \sim T^{-4}$ cannot be validated because of increasing contributions from surface phonons. The observed $\tau_{e-ph} \sim T^{-3}$ dependence coinciding with the temperature dependence for pure bulk metal is likely to be accidental. A partial reduction of phonon dimensionality, when temperature decreases, must suppress the predicted power exponent below 4. This is one of the factors contributing to the observed lower exponent. It is not the only factor. The equilibrium phonon distribution is very broad. Therefore, the substantial fraction of smaller wavelength phonons satisfying the condition $ql \gg 1$, will behave as 3D phonons in the pure crystal with contributions to $\tau_{e-ph}(T)$ scaling as $T^{-3}$.

An important feature that follows from our analysis and simulation of experimental data is the substantial deviation of lattice heat capacitance from the Debye law. Thin films of WSi used for SNSPDs are amorphous. Besides, they are grown on amorphous $SiO_2$ substrate. The characteristic feature of amorphous solids is the excess lattice heat capacity coinciding with the heat conductivity plateau at intermediate temperatures above 1 K. This is a strong indication of the important role of the two level systems and local oscillators in electron energy relaxation in WSi. However, this requires a special study of vibrational properties of amorphous WSi at low temperatures and detailed work on electron-phonon interaction, which goes far beyond the current work.

## V. CONCLUSION

We have shown that the conventional two-temperature model becomes inadequate for the description of the energy exchange between acoustically soft thin metal film and acoustically rigid substrate. We found that the rate of the energy exchange in this case may substantially slow down as a result of the internal phonon bottleneck effect. This effect is important for both 3D and 2D phonon systems and originates from the splitting of the phonon spectrum into sets of non-escaping and escaping modes. The electrons and non-escaping phonons form a unified subsystem, which is cooled down only via the interactions with the subsystem of escaping phonons, either via a direct phonon conversion due to the elastic scattering or via an indirect sequential interaction with the electronic system. The identified regime is qualitatively different both from the electron heating and the bolometric regime. In contrast to pure electron heating, the energy relaxation time is not the electron-phonon time, and is enlarged compared to the former due to the fact that only the interaction with the escaping phonon mode remains efficient for energy relaxation. In contrast to the bolometric regime, the energy relaxation time is not set by heat conductance of the film/substrate interface and, in particular, should be independent of the thickness for sufficiently thin films. We characterized WSi films as a material for developing SNSPD and obtained the magnitude of electron-phonon relaxation time to be in the range $\tau_{e-ph}$ ~ 150-200 ps at 3.4 K in line with the estimates derived by other experimental techniques. Finally, we found that experimental data can only be simulated under the assumption of lattice heat capacity in WSi being substantially increased compared to the Debye model at a low temperature. We show that the excess phonon density of states in amorphous WSi together with the effects of phonon quantization in thin film is likely to be responsible for the observed $\sim T^{-3}$ temperature dependence of electron-phonon relaxation time.

## ACKNOWLEDGMENTS

The work was supported by the Ministry of Education and Science of the Russian Federation, contract No. 14.B25.31.0007, by Russian Foundation for Basic Researches, grant No. 16-29-11779 ofi_m, and also was implemented in the framework of the Basic Research Program at the National Research University Higher School of Economics (HSE) in 2016. M.S. acknowledge the support by Russian Foundation for Basic Researches (grant No. 16-32-00653 mol_a), A.S. acknowledges the support by the State task for institutions of higher education (project No. 2575), G.G. acknowledges the support by the State task for institutions of higher education (project No. XXXX), A. Kozorezov gratefully acknowledges financial support from the Engineering and Physical Sciences Research Council (UK) and DARPA (USA). The authors thank T. M. Klapwijk, A.D. Semenov and D. Yu. Vodolazov for discussion of the results.